\documentclass[prd,preprint,superscriptaddress,preprintnumbers,eqsecnum,showpacs,nofootinbib,nobibnotes]{revtex4-1}

\usepackage{amsmath}
\usepackage{amssymb}
\usepackage{graphicx}
\usepackage{color}
\usepackage{xspace}
\usepackage[utf8]{inputenc}

\usepackage[mathscr,scaled=1.15]{urwchancal}
\DeclareFontFamily{OT1}{pzc}{}
\DeclareFontShape{OT1}{pzc}{m}{it}%
{<-> s * [1.15] pzcmi7t}{}
\DeclareMathAlphabet{\mathpzc}{OT1}{pzc}{m}{it}

\definecolor{purple}{rgb}{0.5,0,0.5}
\definecolor{blue}{rgb}{0.0,0,0.9}
\definecolor{prdblue}{rgb}{0.133,0.118,0.498}

\def\s#1{{\scriptscriptstyle #1}}

\def\noeq#1{(\ref{#1})}
\def\1eq#1{Eq.~(\ref{#1})}

\def\2eqs#1#2{Eqs.~(\ref{#1}) and~(\ref{#2})}
\def\3eqs#1#2#3{Eqs.~(\ref{#1}),~(\ref{#2}) and~(\ref{#3})}

\def\fig#1{Fig.~\ref{#1}}

\def\diff{{\rm d}}

\def\ie{{\it i.e.}, }

\def\n#1{({\it #1})}

\def\g{\widetilde\Gamma^{ \bf np}}

\def\g{\widetilde\Gamma^{ \bf np}}

\def\h{m^2}
\def\N{{c}}

\def\aBSE{\alpha_s^\s{\mathrm{BSE}}}
\def\aDSE{\alpha_s^\s{\mathrm{SDE}}}
\def\lBSE{\lambda_\s{\mathrm{BSE}}}
\def\lDSE{\lambda_\s{\mathrm{SDE}}}
\def\fQ{f}

\begin{document}

\title{Coupled dynamics in gluon mass generation \\ and the impact of the three-gluon vertex}

\author{Daniele Binosi}
\affiliation{European Centre for Theoretical Studies in Nuclear Physics and Related Areas (ECT*) and Fondazione Bruno Kessler, \\Villa Tambosi, Strada delle Tabarelle 286, I-38123 Villazzano (TN)  Italy}
\author{Joannis Papavassiliou}
\affiliation{\mbox{Department of Theoretical Physics and IFIC, University of Valencia and CSIC}, E-46100, Valencia, Spain}

\date{11 July 2017}

\begin{abstract}

We present  a detailed  study of the  subtle interplay 
transpiring  at  the   level  of two  integral equations  that  are
instrumental for the  dynamical 
generation of  a
gluon  mass in  pure
Yang-Mills theories.  The main novelty 
is the joint treatment of the 
Schwinger-Dyson  equation governing the infrared  behaviour of
the  gluon propagator  and of the integral equation  that controls  the
formation of massless bound-state excitations, whose inclusion is instrumental 
for obtaining massive solutions
from the former equation.
The self-consistency  of the entire approach imposes  
the requirement of using a single value for the gauge coupling
entering in the two key equations; its 
fulfillment  depends crucially  on the details of the  
three-gluon vertex, which contributes to both of them, but 
with different weight. 
In particular,
the characteristic suppression of this vertex 
at intermediate and low energies
enables the convergence of the iteration procedure to 
a single gauge coupling, 
whose  value is reasonably close to that 
extracted from related lattice simulations.

\end{abstract}

\maketitle

\section{Introduction}
The nonperturbative aspects of the gluon propagator, $\Delta^{ab}_{\mu\nu}(q)$, are considered to be especially relevant for the qualitative and quantitative understanding of a wide range of important physical  phenomena, such as confinement, chiral symmetry breaking, and bound-state formation. A particularly interesting feature, which manifests itself both in the Landau gauge and away from it, is the saturation of its scalar form factor, $\Delta(q^2)$, in the deep infrared (IR), \ie  $\Delta(0) = c_0 >0$. This special behavior, which is believed to be intimately connected with the emergence of a fundamental mass scale, was firmly established in a variety of SU(2)~\cite{Cucchieri:2007md,Cucchieri:2007rg,Cucchieri:2009zt} and SU(3)~\cite{Bowman:2007du,Bogolubsky:2009dc,Oliveira:2009eh,Ayala:2012pb,Bicudo:2015rma} large-volume lattice simulations, and has been extensively studied in the continuum within diverse theoretical frameworks~\cite{Cornwall:1981zr,Lavelle:1991ve,Halzen:1992vd,Philipsen:2001ip,Szczepaniak:2001rg,Aguilar:2004sw,Aguilar:2006gr,Kondo:2006ih,Braun:2007bx,Epple:2007ut,Aguilar:2008xm,Boucaud:2008ky,Dudal:2008sp,Fischer:2008uz,Aguilar:2009nf,RodriguezQuintero:2010wy,Campagnari:2010wc,Tissier:2010ts,Kondo:2010ts,Pennington:2011xs,Watson:2011kv,Kondo:2011ab,Serreau:2012cg,Strauss:2012dg,Cloet:2013jya,Siringo:2014lva,Binosi:2014aea,Aguilar:2015nqa,Huber:2015ria,Capri:2015ixa,Binosi:2016nme,Glazek:2017rwe,Gao:2017uox}.

In one of these approaches, the  Schwinger-Dyson equation (SDE) that controls the evolution of the gluon propagator has been shown to yield an infrared finite (``massive'') solution. We emphasize that the relevant SDE was formulated within the framework developed through the merging of the pinch-technique (PT)~\cite{Cornwall:1981zr,Cornwall:1989gv,Pilaftsis:1996fh,Binosi:2002ft,Binosi:2003rr,Binosi:2009qm} with the  background-field method (BFM)~\cite{Abbott:1980hw}, to be referred to as ``PT-BFM scheme''~\cite{Aguilar:2006gr,Binosi:2007pi,Binosi:2008qk}. Inherent to this scheme is the distinction between background ($B$) and quantum ($Q$) gluons, and the proliferation of the possible Green's functions that one may form with them. Particularly relevant for what follows is the distinction between the $QQ$ and $QB$ gluon self-energies, and the $Q^3$ and $BQ^2$ three-gluon vertices, to be denoted by $\Gamma$ and $\widetilde \Gamma$, respectively.

An indispensable ingredient for the realization of this scenario is the presence of massless poles of the type $1/q^2$ in the vertices with one $B$ leg, which enter into the $QB$ gluon self-energy~\cite{Aguilar:2011xe,Ibanez:2012zk,Aguilar:2016vin,Aguilar:2016ock} and implement the well-known Schwinger mechanism for gauge-boson mass generation~\cite{Schwinger:1962tn,Schwinger:1962tp,Jackiw:1973tr,Smit:1974je,Eichten:1974et,Poggio:1974qs}. The origin of these poles is dynamical, owing to the formation of {\it colored} bound-state excitations, which are massless  due to the strong binding induced by the Yang-Mills interactions. The  integral equations that govern their formation constitute a system of homogeneous linear Bethe-Salpeter equations (BSEs), which determines the derivatives of the corresponding ``bound-state wave functions''. In the present work we will simplify the degree of complexity by restricting the possibility of pole formation {\it only} in $\widetilde \Gamma$, thus reducing the aforementioned system into a {\it single} BSE, which determines the corresponding derivative, to be denoted by~$\widetilde{C}'_{1}(k^2)$.

Evidently, the self-consistent implementation of the dynamical picture described above hinges on the subtle interplay between the BSE and SDE, and the compatibility of the various field-theoretic ingredients that enter in them. The purpose of the present work is to focus on a particularly pivotal aspect of this interplay, and elucidate the decisive impact not only of $\widetilde \Gamma$, whose $1/q^2$ pole enforces the desired infrared finiteness of $\Delta(q^2)$, but especially of $\Gamma$, whose infrared structure affects both the kernel of the BSE and a crucial two-loop component of the SDE. 

In order to appreciate the circumstances described above in some detail, let us first observe that the pole BSE and the gluon SDE are tightly intertwined
mainly because \n{i} the SDE expresses the value of $\Delta^{-1}(0)$ as an integral that involves $\widetilde{C}'_{1}(k^2)$~\cite{Aguilar:2016vin}, while, at the same time, \n{ii} $\widetilde{C}'_{1}(k^2)$ is known to be proportional to $\diff m^2(k^2)/\diff k^2$~\cite{Aguilar:2011xe}. Thus, once obtained from the BSE, it provides, upon integration, the running gluon mass $m^2(k^2)$, a notion that dates back to the pioneering work of~\cite{Cornwall:1981zr}. This dual role of $\widetilde{C}'_{1}(k^2)$, coupled to the obvious requirement that $\Delta^{-1}(0) = m^2(0)$, imposes finally a stringent constraint on the strong coupling $\alpha_s = g^2/4 \pi$; specifically, the value of $\alpha_s$ used in \n{i}, to be denoted by $\aDSE$, ought to coincide with that employed in \n{ii}, to be denoted by  $\aBSE$.

As advocated above, the nonperturbative behavior of the vertex  $\Gamma$ becomes relevant when trying to enforce the equality $\aDSE=\aBSE$. 
Note in particular that \n{a} $\Gamma$ enters {\it linearly} in the SDE-derived expression that determines the value of $\Delta^{-1}(0)$  and  {\it quadratically} in the kernel of the BSE, rendering it renormalization group invariant (RGI), 
and \n{b}  below 1 {\rm GeV} the vertex $\Gamma$ 
is suppressed with respect 
to its tree-level value,
reversing its sign around 100 {\rm MeV}, and finally diverging logarithmically at the origin. 

It turns out that, when the tree-level expression of $\Gamma$ is used in the evaluation of \n{i} and \n{ii}, the resulting values for $\aDSE$ and $\aBSE$ {\it do not} coincide. Instead, if one employs a standard nonperturbative Ansatz for $\Gamma$, which encodes the features mentioned in \n{b}, one finds that, indeed, $\aDSE=\aBSE$. The common value is given by  $\alpha_s =0.45$, when the momentum subtraction (MOM) renormalization is implemented at $\mu=4.3 {\rm GeV}$. This particular value for $\alpha_s$ is  to be contrasted with the one obtained (for the same $\mu$) from the lattice simulation of the three-gluon vertex $\Gamma$ in~\cite{Athenodorou:2016oyh}, namely $\alpha_s =0.32$. This discrepancy appears to be more than acceptable given the approximations implemented when deriving both the SDE and the BSE, and, in particular, the simplifications applied in the renormalization of the former, and the truncations imposed when constructing the kernel of the latter.

\section{Schwinger mechanism and vertices with massless poles}
\begin{figure}[!t]
    \centering
    \mbox{}\hspace{-0.8cm}
    \includegraphics[scale=0.6]{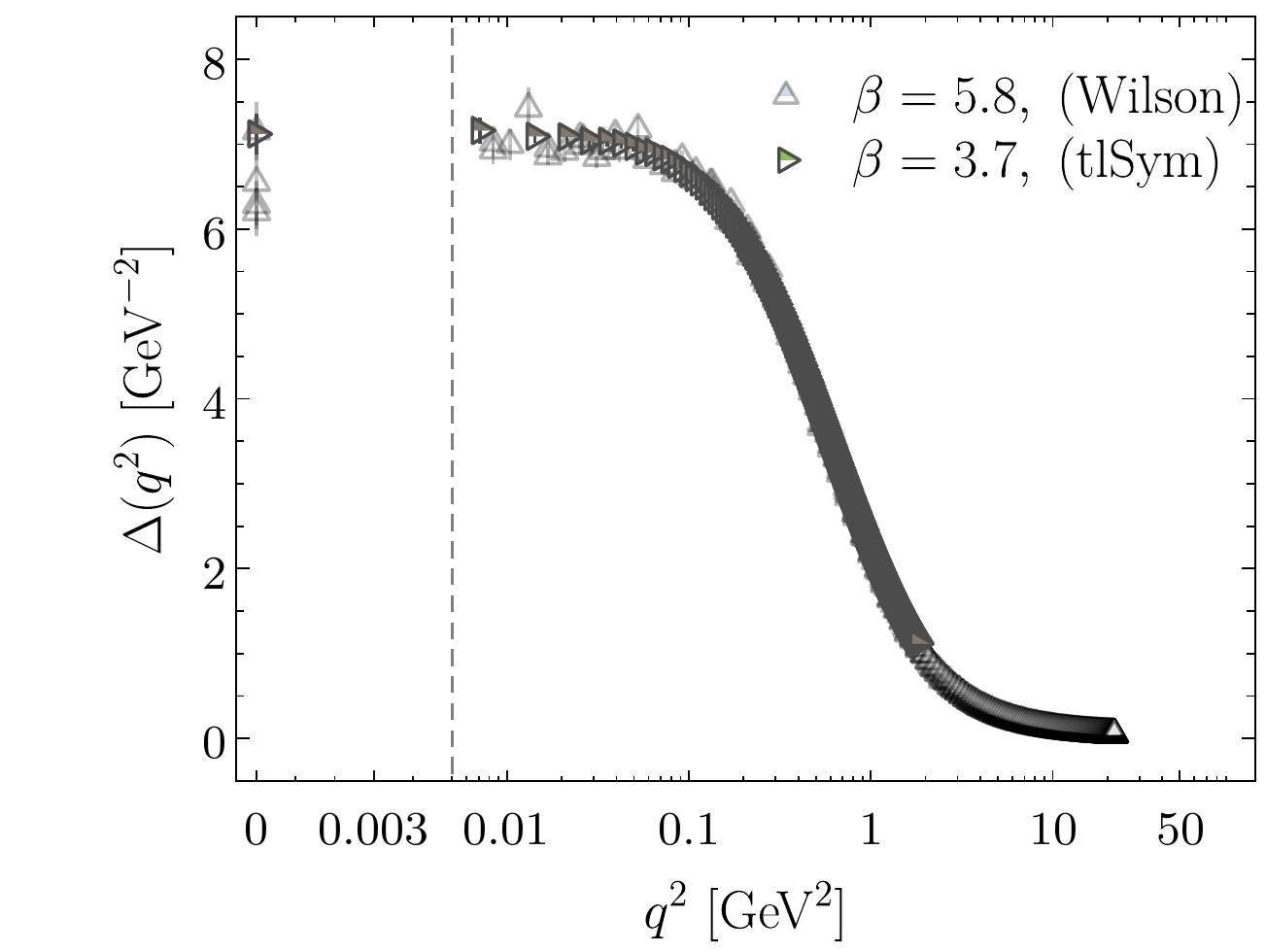}
    \caption{\label{fig:glprop}Lattice data for the quenched Landau gauge gluon propagator  obtained from a tree-level Symanzik (tlSym) improved gauge action~\cite{Athenodorou:2016oyh,Boucaud:2017obn} (calibrated following the procedure described in~\cite{Boucaud:2017ksi}), compared with the corresponding data obtained from a Wilson gauge action~\cite{Bogolubsky:2009dc}. The momentum axis is linear on the left of the vertical dashed line and logarithmic on the right, an artifice that clearly exposes the existence of a saturation point at IR momenta.}
\end{figure}

Throughout this work, we consider a SU(3) pure Yang-Mills theory (no dynamical quarks).
In the Landau gauge, the gluon propagator \mbox{$\Delta^{ab}_{\mu\nu}(q)= 
\delta^{ab}\Delta_{\mu\nu}(q)$} has the form  
\begin{align}
\Delta_{\mu\nu}(q) = -i\Delta(q^2)P_{\mu\nu}(q); \qquad P_{\mu\nu}(q) = g_{\mu\nu}-\frac{q_\mu q_\nu}{q^2},
\label{QQprop}
\end{align}
where $\Delta(q^2)$ is related to the form factor of the gluon 
self-energy \mbox{$\Pi_{\mu\nu}(q) = P_{\mu\nu}(q) \Pi (q^2)$}, through \mbox{$\Delta^{-1}(q^2) = q^2 + i \Pi (q^2)$}. Lattice data for this (quenched) quantity, renormalized at~\mbox{$\mu=4.3$ GeV}, are shown in~\fig{fig:glprop},
and will serve as the main input in the ensuing analysis.
In addition, the ghost propagator 
\mbox{$D^{ab}(q^2) = i\delta^{ab} D(q^2)$} furnishes  
the dressing function, $F(q^2)$, defined as 
\mbox{$F(q^2) = q^2 D(q^2)$}; in the Landau gauge (again at  $\mu=4.3$ GeV), $F(0)\approx2.9$ 

\begin{figure*}[!t]
    \centering
    \includegraphics[scale=0.5]{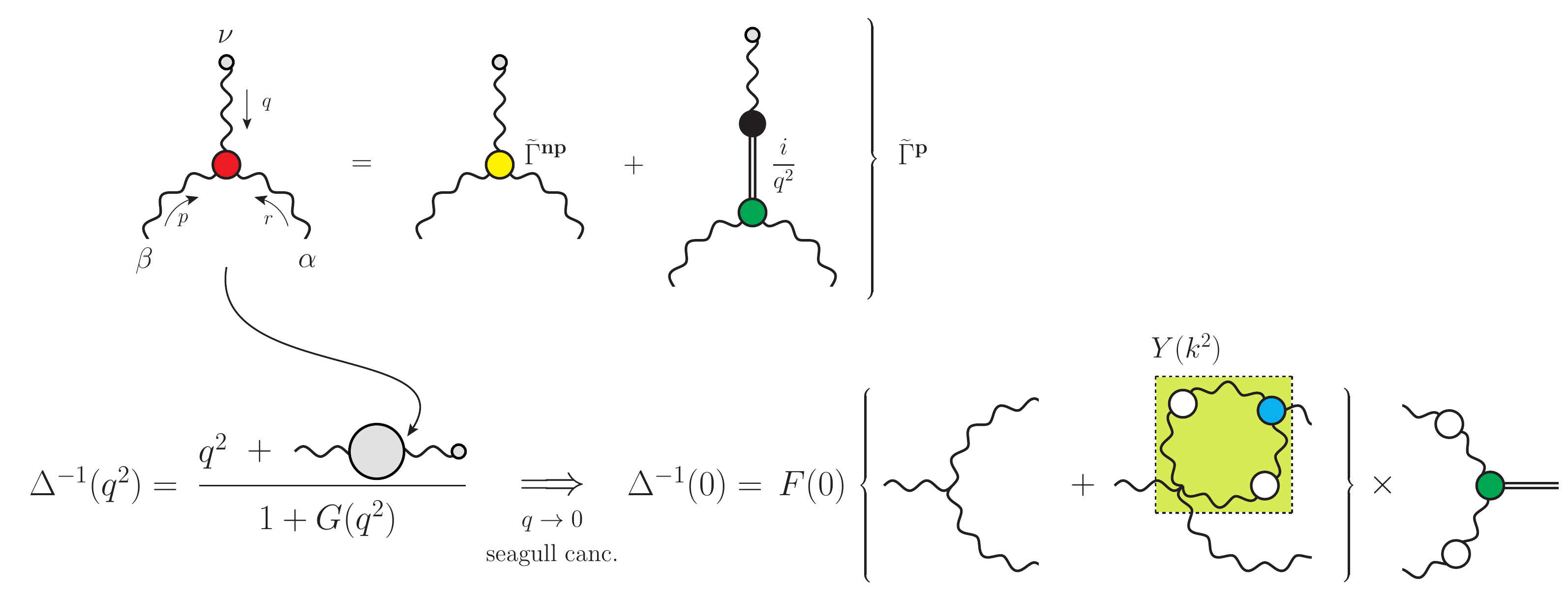}
    \caption{\label{fig:procedure}The procedure leading to the dynamical gluon mass generation within the PT-BFM framework.}
\end{figure*}

In the PT-BFM framework, the SDE of 
$\Delta(q^2)$ is expressed in terms
of the $QB$ self-energy $\widetilde\Pi_{\mu\nu}(q)$, namely  
(see~\fig{fig:procedure}) 
\begin{align}
	\Delta^{-1}(q^2)P_{\mu\nu}(q) &= \frac{q^2P_{\mu\nu}(q)  + i
	\widetilde{\Pi}_{\mu\nu}(q)}{1 + G(q^2)}, 
\label{glSDE1}
\end{align}
where $G(q^2)$ is the $g_{\mu\nu}$ component of a special two-point function~\cite{Binosi:2002ez}.
In the Landau gauge only, the important relation 
\mbox{$1+G(0) = F^{-1}(0)$}  holds exactly~\cite{Grassi:2004yq,Aguilar:2009nf}.

The main advantage of 
expressing the gluon SDE in terms of $\widetilde\Pi_{\mu\nu}(q)$ 
rather than $\Pi_{\mu\nu}(q)$ 
arises from the fact that, 
when contracted from the side of the $B$-gluon, each fully dressed vertex 
satisfies a linear (Abelian-like) Slavnov-Taylor identity (STI). 
In particular, the $BQ^2$ vertex $\widetilde{\Gamma}_{\mu\alpha\beta}$ 
satisfies (color omitted)
\begin{align}
	q^\mu \widetilde{\Gamma}_{\mu\alpha\beta}(q,r,p) = i\Delta_{\alpha\beta}^{-1}(r) - i\Delta_{\alpha\beta}^{-1}(p), 
\label{AbWIthree}
\end{align}

Let us now turn to the infrared dynamics described by \1eq{glSDE1}. 
As has been shown recently~\cite{Aguilar:2016vin}, if the 
vertices carrying the $B$ leg do not contain massless poles of the type 
$1/q^2$, then the $\Delta(q^2)$ 
governed by \1eq{glSDE1} remains rigorously  massless. 
The demonstration relies on the subtle interplay between the Ward-Takahashi identities (WTIs), satisfied by the vertices as $q\to 0$, and an integral relation known as the ``seagull identity''~\cite{Aguilar:2009ke,Aguilar:2016vin}. The basic steps that lead to this result may be exemplified in terms of the $BQ^2$ vertex $\widetilde{\Gamma}_{\mu\alpha\beta}$;
the inclusion of the remaining vertices is conceptually straightforward~\cite{Aguilar:2016vin}.

To that end, 
consider the limit of the STI~\noeq{AbWIthree} as $q\to 0$, 
assuming that $\widetilde{\Gamma}_{\mu\alpha\beta}$ 
does not contain $1/q^2$ terms. Then, the Taylor expansion of both sides 
generates the corresponding WTI  
\begin{align}
    \widetilde{\Gamma}_{\mu\alpha\beta}(0,r,-r) = -i \frac{\partial }{\partial r^\mu}\Delta^{-1}_{\alpha\beta}(r),
\label{BQ2-id}	
\end{align}
which, when used in the evaluation of the gluon SDE, yields 
\begin{align}
    \Delta^{-1}(0) & = \underbrace{\int_k\frac{\partial}{\partial k_\mu}{\cal F}_\mu(k) =0}_{\rm seagull \,\, identity};&{\cal F}_\mu(k) =k_\mu{\cal F}(k^2)
    \label{seag1}
\end{align}
where ${\cal F}(k^2) = \Delta(k^2) [c_1 + c_2 Y(k^2)]$, 
with $c_1,c_2 \neq 0$, and 
\begin{align}
    Y(k^2)&= \frac{1}{(d-1)}\frac{k_\alpha}{k^2}\! \int_\ell\!\Delta^{\alpha\rho}(\ell)\Delta^{\beta\sigma}(\ell+k)\Gamma_{\sigma\rho\beta}(-\ell-k,\ell,k).
    \label{defY}    
\end{align}
Note that  
we have introduced the  dimensional regularization integral measure $\int_{k}\equiv\frac{\mu^{\epsilon}}{(2\pi)^{d}}\!\int\!\mathrm{d}^d k$, with $d=4-\epsilon$ the space-time dimension, and $\mu$  the 't Hooft mass scale.

\begin{figure}[!t]
    \centering
    \includegraphics[scale=0.5]{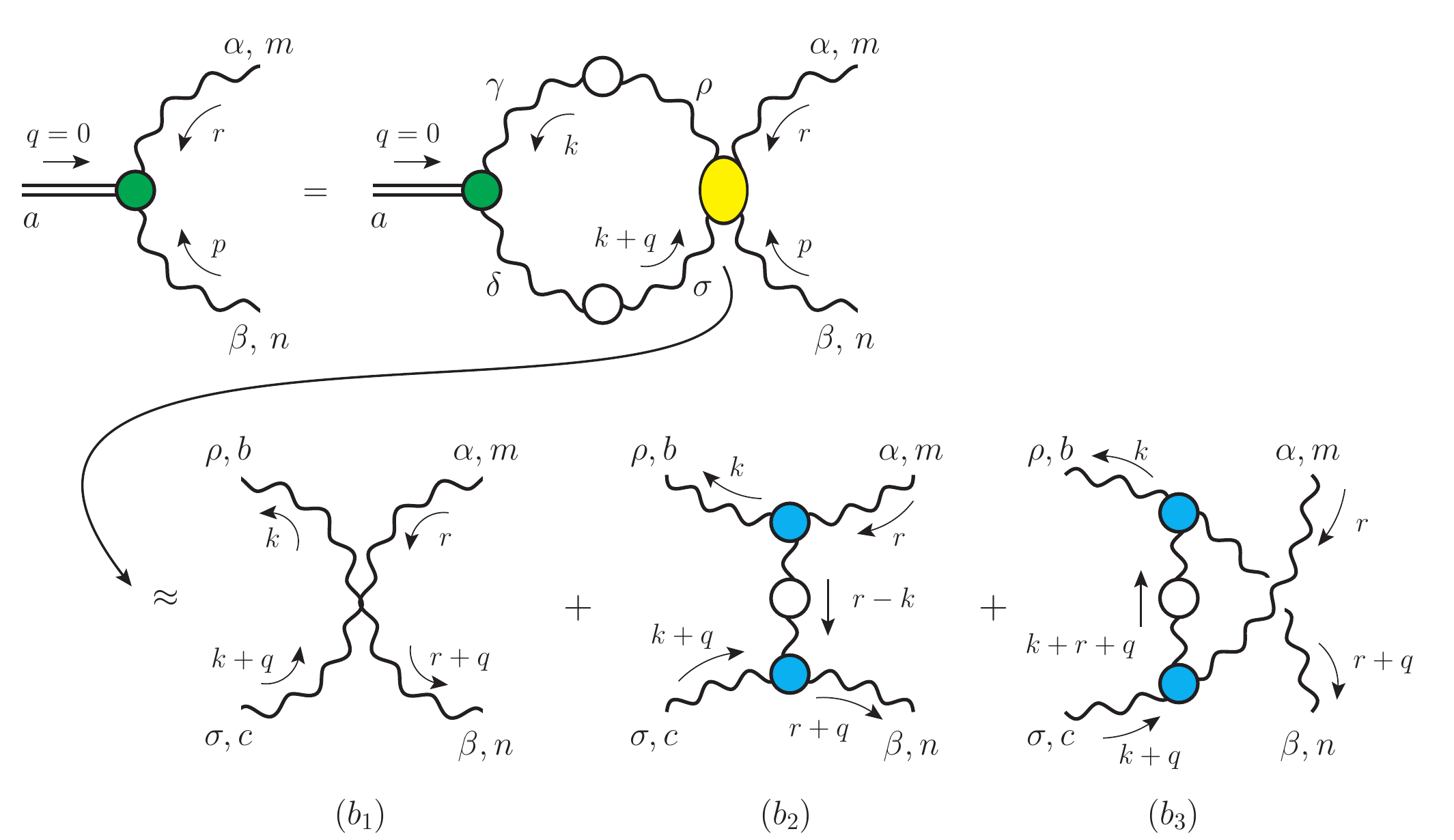}
    \caption{\label{fig:bse-4g}The BSE satisfied by the bound-state wave function $\widetilde{C}_{\alpha\beta}$ (upper line) and the simplified four gluon kernel used.}
\end{figure}

In order to circumvent the result of \1eq{seag1}, 
one must allow $\widetilde{\Gamma}_{\mu\alpha\beta}$ to contain 
longitudinally coupled $1/q^2$ poles; 
their inclusion, 
in turn, triggers the Schwinger mechanism~\cite{Schwinger:1962tn,Schwinger:1962tp}, 
finally enabling the generation of a gauge boson mass~\cite{Jackiw:1973tr,Smit:1974je,Eichten:1974et,Poggio:1974qs}. More specifically,  we have 
\begin{align}
    \widetilde{\Gamma}_{\mu\alpha\beta}(q,r,p) = \g_{\mu\alpha\beta}(q,r,p) + \frac{q_\mu}{q^2}\widetilde{C}_{\alpha\beta}(q,r,p),
\label{GnpGp}
\end{align}
where the superscript ``np'' indicates the ``no-pole'' part, and~$\widetilde{C}_{\alpha\beta}$ is the aforementioned bound-state wave function.
Evidently, the Bose-symmetry of the vertex under the exchange 
 \mbox{$(\alpha$, $r$) $\leftrightarrow$ ($\beta$, $p$)} imposes  the relation 
 $\widetilde{C}_{\alpha\beta}(0,r,-r) =0$. 
 
Next, in order to preserve the BRST symmetry of the theory, 
we demand that all STIs maintain their exact form in the presence of these poles; therefore, \1eq{AbWIthree} will now read  
\begin{align}
	q^\mu \g_{\mu\alpha\beta}(q,r,p) + \widetilde{C}_{\alpha\beta}(q,r,p) = i\Delta_{\alpha\beta}^{-1}(r) - i\Delta_{\alpha\beta}^{-1}(p).
\label{STIwP}
\end{align}
Taking the limit of~\1eq{STIwP} as $q\to 0$ 
and matching the lowest order terms in $q$, 
the corresponding WTI becomes  
\begin{align}
	\g_{\mu\alpha\beta}(0,r,-r) = -i\frac{\partial}{\partial r^\mu}\Delta^{-1}_{\alpha\beta}(r) - \left\lbrace\frac{\partial}{\partial q^\mu}\widetilde{C}_{\alpha\beta}(q,r,-r-q)\right\rbrace_{q=0}.
\label{WIwithpole}
\end{align}

The presence of the second term on the r.h.s. 
of \1eq{WIwithpole}
has far-reaching consequences for the infrared behavior of $\Delta$. Specifically, a repetition of the steps leading to \1eq{seag1} reveals that, whereas the first term on the r.h.s. of \1eq{WIwithpole} reproduces again \1eq{seag1} (and its contribution thus vanishes), the second term survives, giving 
\begin{align}
    \Delta^{-1}(0)&=\frac32g^2C_AF(0)\int_k k^2 \Delta^2(k^2)\left[1-\frac32g^2C_AY(k^2)\right]\widetilde{C}_1'(k^2),
    \label{DSEmass}
\end{align}
where $\widetilde{C}_1(q,-k-q,k)$ is the form factor of $g_{\alpha\beta}$  
in the tensorial decomposition of $\widetilde{C}_{\alpha\beta}$, 
\begin{align}
\widetilde{C}_1^{\prime}(k^2)=\underset{q\rightarrow 0}{\lim}\left\lbrace\frac{\partial \widetilde{C}_1(q,k,-k-q)}{\partial (k+q)^2}\right\rbrace\,,
\end{align}
and $C_A$ is the Casimir eigenvalue of the adjoint representation [$N$ for SU($N$)]. 
Note that 
the one- and two-loop dressed contributions 
enter into the mass 
condition~\noeq{DSEmass} with a {\it different relative sign}, 
a fact that is crucial for 
the ensuing analysis. 

\section{BSE for the massless bound-states}
The dynamical equation that governs $\widetilde{C}_{1}(k^2)$ may be derived from the SDE satisfied  $\widetilde{\Gamma}_{\mu\alpha\beta}(q,r,p)$, as $q \to 0$. In this limit, the derivative term becomes the leading contribution, given that  $\widetilde{C}_{\alpha\beta}(0,r,-r)=0$, 
and the resulting homogeneous equation assumes the form of a BSE (see~\fig{fig:bse-4g}), given by~\cite{Aguilar:2016ock}
\begin{align}
    f^{amn}\lim_{q\to 0} \widetilde{C}_{\alpha\beta}(q,r,p)
    &= f^{abc}\lim_{q\to 0} \Bigg\{ \int_k \widetilde{C}_{\gamma\delta}(q,k,-k-q)\Delta^{\gamma\rho}(k)\nonumber \\
    &\times\Delta^{\delta\sigma}(k+q){\cal K}_{\rho\alpha\beta\sigma}^{bmnc}(-k,r,p,k+q)\Bigg\}.
\label{BSEq}
\end{align}
To proceed further, we will approximate the four-gluon BS kernel ${\cal K}$ by the  lowest-order set of diagrams appearing in its skeleton expansion, given by the 
diagrams $(b_1)$, $(b_2)$, and $(b_3)$,
shown in the second line 
of \fig{fig:bse-4g}. It turns out that, 
if we use the tree-level four-gluon vertex 
in the evaluation of $(b_1)$,
its contribution 
in the above kinematic limit 
vanishes.
Diagrams $(b_2)$ and $(b_3)$, which carry a statistical factor of 1/2, 
are considered to contain fully dressed 
gluon propagators and 
three gluon vertices $\Gamma$
(note that all gluons are of the $Q$-type). 
As a consequence, the resulting BSE 
does not depend on the value of the 
MOM subtraction point $\mu$, 
because the two graphs composing its 
kernel may be written as  
the ``square'' of the formally 
RGI combination 
\begin{align}
{\cal R}^{\mu\alpha\beta}(q_1,q_2,q_3) =  g \Delta(q_1)\Delta^{1/2}(q_2) \Gamma^{\mu\alpha\beta}(q_1,q_2,q_3),
\label{RGIcom}
\end{align}
namely, setting $q=0$,  
\mbox{$(b_2) \sim {\cal R}(-k,k-r,r)
{\cal R}(k,r-k,-r)$} and 
$(b_3) \sim {\cal R}(-k,k+r,-r)$
${\cal R}(k,-r-k,r)$.

The vertex $\Gamma$ 
contains 14 form factors~\cite{Ball:1980ax}, whose nonperturbative structure, 
albeit subject of various studies~\cite{Alkofer:2008dt,Tissier:2011ey,Pelaez:2013cpa,Aguilar:2013vaa, Blum:2014gna,Eichmann:2014xya,Williams:2015cvx,Cyrol:2016tym},  
is only partially known. Therefore, 
for the purposes of the present work,
we will consider the simple Ansatz
\begin{align}
	\Gamma_{\mu\alpha\beta}(q,r,p)=
	\fQ (r)\Gamma^{(0)}_{\mu\alpha\beta}(q,r,p),
	\label{vertf}
\end{align} 
where $\Gamma^{(0)}$ is the 
standard tree-level expression of the vertex, and the form factor $\fQ (r)$ is considered to be a function of a single kinematic variable. Then, using \1eq{vertf} into \1eq{BSEq}, we arrive at the final equation 
\begin{align}
    &\widetilde{C}_1'(q^2)=\frac{8\pi}3\alpha_s C_A\!\!\!\int_k\!\widetilde{C}_1'(k^2)\frac{(q\!\cdot\!k)[q^2k^2-(q\!\cdot\! k)^2]}{q^4k^2(k+q)^2}\Delta^2(k)\Delta(k+q)\nonumber\\
    &\times \fQ^2(k+r)\left[8q^2k^2 + 6(q\!\cdot\!k)(q^2+k^2)+3(q^4+k^4)+(q\!\cdot\! k)^2\right].
    \label{masslessBSE}
\end{align}

\begin{figure}[!t]
    \centering
    \mbox{}\hspace{-0.8cm}
    \includegraphics[scale=0.6]{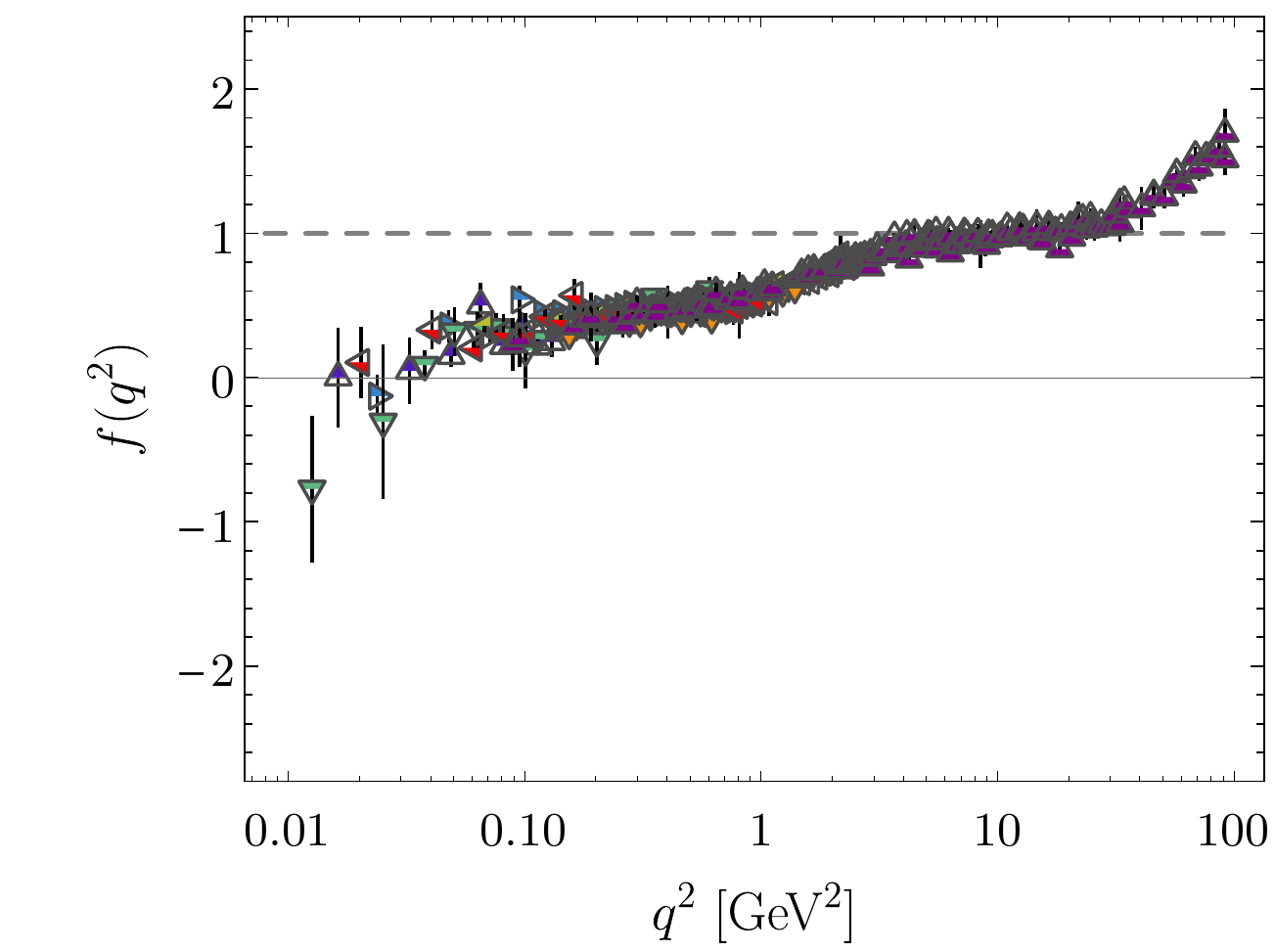}
    \caption{\label{fig:fQ3}Compilation of SU(3) lattice data (evaluated with various $\beta$, volumes and actions) for the form factor $\fQ $ in the symmetric configuration~\cite{Athenodorou:2016oyh,Boucaud:2017obn}.}
\end{figure}

The functional form we will employ
for $\fQ (r)$ is motivated by a considerable number 
of lattice simulations and studies in the continuum. 
In particular, for 
certain characteristic kinematic configurations (such as the symmetric and the 
soft gluon limits), the vertex is suppressed 
with respect to its tree-level value, reverses its sign for 
relatively small momenta (an effect known as ``zero crossing''), 
and finally diverges at the origin~\cite{Alkofer:2008dt,Tissier:2011ey,Pelaez:2013cpa,Aguilar:2013vaa, Blum:2014gna,Eichmann:2014xya,Williams:2015cvx,Cyrol:2016tym}. 
The reason for this particular behavior may be traced back to 
the delicate balance between contributions originating from 
gluon loops, which are ``protected'' by the 
corresponding gluon mass, and the ``unprotected'' logarithms 
coming from the ghost loops that contain massless ghosts.
Early lattice indication for a zero crossing in SU(2) Yang-Mills theories can be found in~\cite{Cucchieri:2006tf,Cucchieri:2008qm}, whereas the effect has been recently confirmed to be present also in the case of SU(3) theories~\cite{Athenodorou:2016oyh,Duarte:2016ieu,Boucaud:2017obn}. A compilation of the lattice data of~\cite{Athenodorou:2016oyh,Boucaud:2017obn}, properly normalized by dividing out the coupling [$g=2$ at $\mu=4.3$ GeV for the set at hand, corresponding to $\alpha_s=0.32$], is shown in~\fig{fig:fQ3}.

\section{Running gluon mass from the BSE}
In the absence of poles, the validity of  
\1eq{seag1} suggests that 
$\Delta^{-1}(q^2)=  q^2 J(q^2)$, where 
the function $J(q^2)$ 
captures the perturbative contributions
and diverges as $\ln q^2$ at the origin. 
Instead, the infrared saturation 
of $\Delta^{-1}(q^2)$  
motivates the physical parametrization  
\begin{align}
    \Delta^{-1}(q^2)& =  q^2 J(q^2) + \h (q^2),
\label{DQH}
\end{align}
with $\h (0)\neq 0$. Note that 
$J(q^2)$ is also affected 
by the presence of the mass, since    
most of its logarithms are now 
``protected''.

If we now introduce \1eq{DQH} 
in the rhs of \1eq{STIwP},  
it is natural to associate the 
$J$ terms with the $q^\mu\g_{\mu\alpha\beta}$
on the l.h.s, 
and, correspondingly, 
\begin{align}
	\widetilde{C}_{\alpha\beta}(q,r,p) &= \h(p^2) P_{\alpha\beta}(p) - \h(r^2) P_{\alpha\beta}(r).
    \label{thectilde}
\end{align}
Focusing on the $g_{\alpha\beta}$ components of \1eq{thectilde},
we obtain 
\begin{align}
\widetilde{C}_1(q,r,p)=\h(r^2)-\h(p^2),  
\end{align}    
which, in the limit $q\to0$, 
leads to the important result~\cite{Aguilar:2011xe} 
\begin{align}  
\widetilde{C}'_1(r^2)=
\frac{\diff m^2(r^2)}{\diff r^2}. 
\end{align}
Then, upon integration, 
\begin{align}
    \h(x)=\Delta^{-1}(0)+\int_0^{x}\!\!\diff y\,\widetilde{C}'_1(y),
    \label{C1vsmass}
\end{align}
where $x=q^2$ and $y=r^2$. \1eq{C1vsmass} establishes thus a possible link between the solution of the BSE~\noeq{masslessBSE} and what has been identified in the literature with the dynamically generated gluon mass~\cite{Binosi:2012sj}. However, in order for the quantity $\h(q^2)$ to admit a running mass interpretation in the sense familiar from the quark case, it needs to: \n{i} be a monotonically decreasing function of $q^2$; \n{ii} vanish in the UV, \ie satisfy $\h(\infty)=0$. 

\begin{figure*}[!t]
    \centering
    \mbox{}\hspace{-.8cm}
    \includegraphics[scale=0.6]{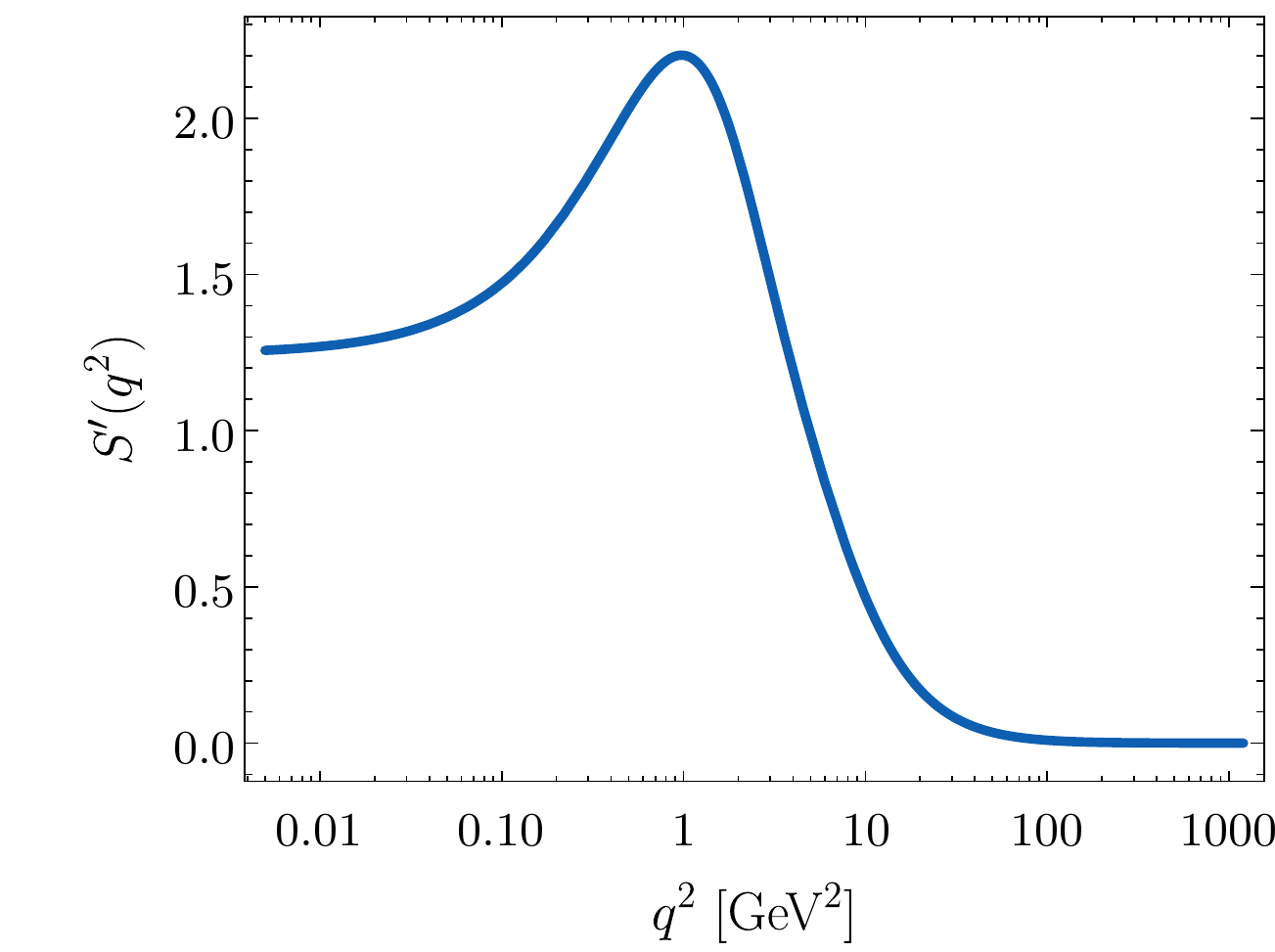}\hspace{.3cm}
    \includegraphics[scale=0.6]{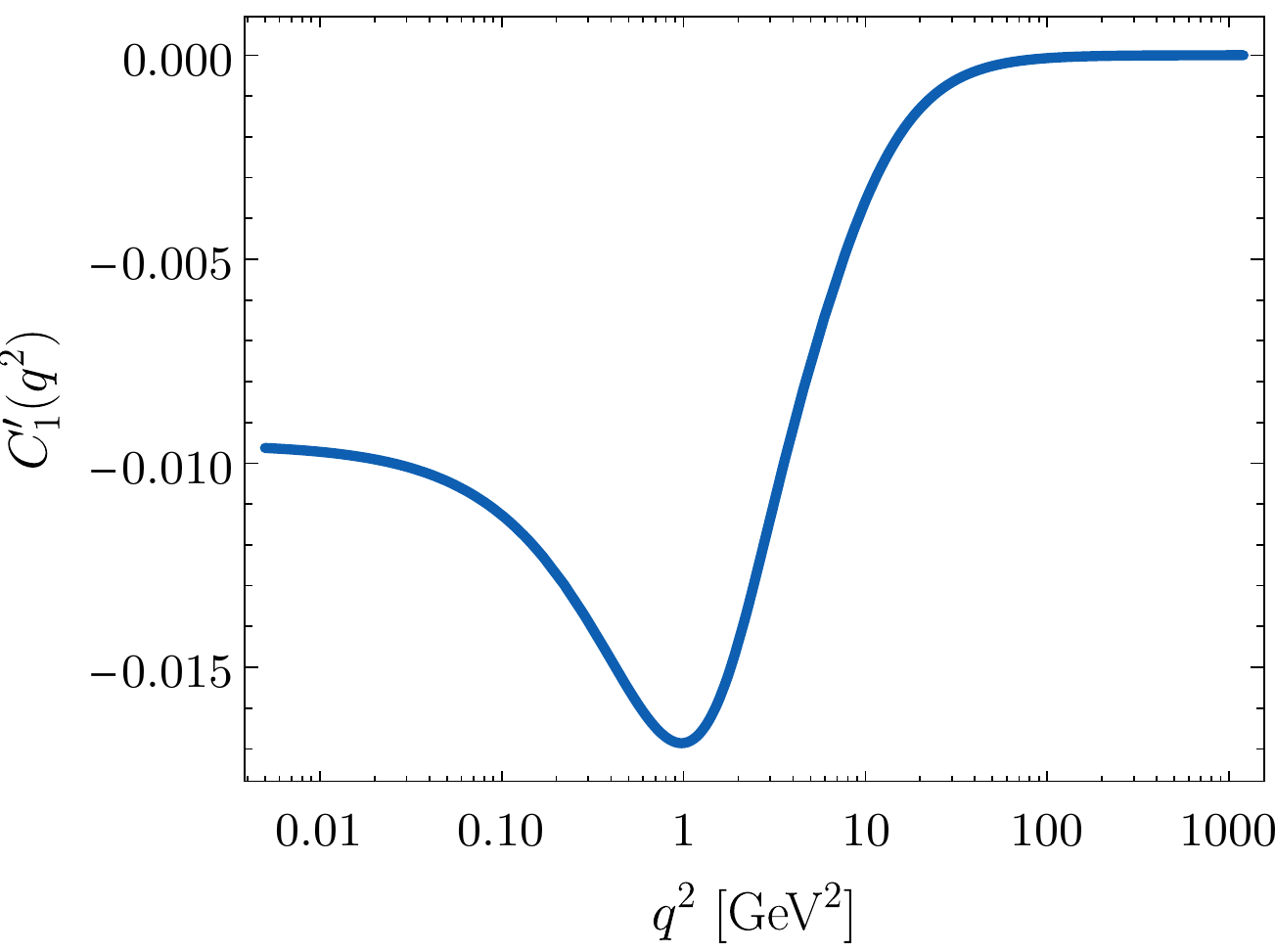}
    \includegraphics[scale=0.6]{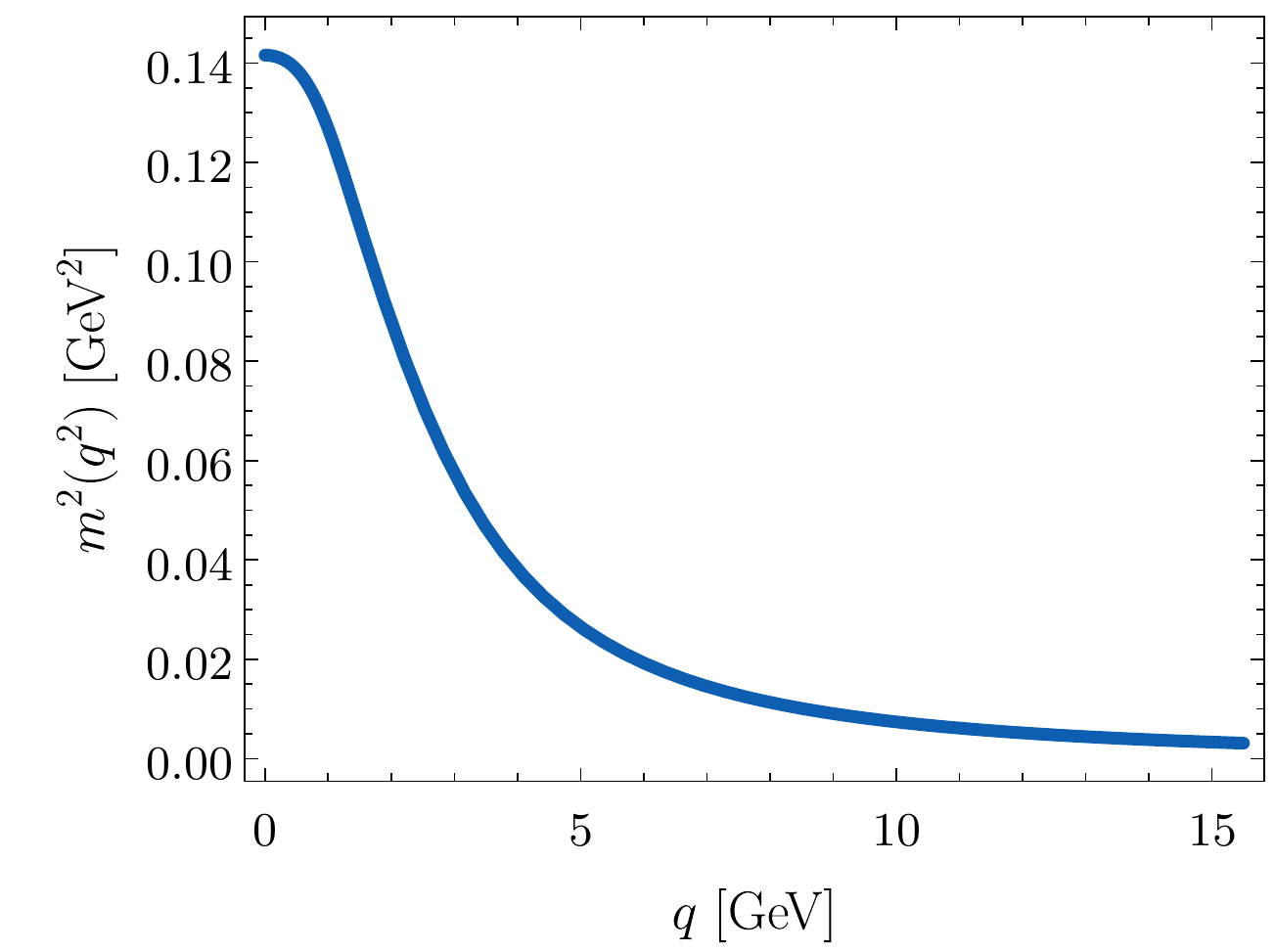}
    \caption{\label{fig:Sprime} From the left-top clockwise: A typical solutions of the BSE~\noeq{masslessBSE}, the corresponding normalized solution, and the associated dynamically generated gluon mass.}
\end{figure*}

To explore the implications of these requirements, let $S'$ be a general solution of the BSE~\noeq{masslessBSE} corresponding to a certain (eigen)value of the strong coupling, $\alpha_s=\aBSE$. 
The typical shape of such solutions is shown in~\fig{fig:Sprime}. Then, one has $\widetilde{C}'_1(x)=\N S'(x)$, where $\N$ is a normalization constant 
that needs to be determined. To this end, observe that, with the kernel used, $S'$ is positive definite; then, the requirement of a monotonically decreasing $\h(x)$ forces $\N$ to be negative: $\N=-\vert\N\vert$. 
Furthermore, the condition $\h(\infty)=0$ fixes its modulus, since~\1eq{C1vsmass} implies
\begin{align}
    \Delta^{-1}(0)=\vert\N\vert\int_0^\infty\!\!\diff y\,S'(y), 
\label{fixc}
\end{align}    
so that $\vert\N\vert\ = 0.0076$. Substitution 
of \1eq{fixc} into \1eq{C1vsmass} yields 
\begin{align}    
\h(x)=\vert\N\vert\int_x^\infty\!\!\diff y\,S'(y),
\label{massfromS}
\end{align}
which, upon integration, gives rise to the squared running mass 
shown in~\fig{fig:Sprime}; it may be 
accurately fitted by
\begin{align}
m^2(q^2)=m^2(0)/[1+(q^2/m_1^2)^{1+p}], 
\end{align}
with $m_1=0.36$ GeV and $p=0.1$, in excellent agreement with 
the behavior found in~\cite{Aguilar:2014tka}. 

\section{BSE/SDE consistency condition}
Let us now return to \1eq{DSEmass},
whose derivation was carried 
out before renormalization. 
Its renormalization 
may be carried out by introducing the 
standard renormalization constants 
for the propagators, vertices, and the 
coupling. Then, using the constraints 
that the various STIs impose on these 
constants, all quantities entering into 
\1eq{DSEmass} can be converted into 
renormalizaed ones, and the 
replacement 
\begin{align}
1-\frac32g^2C_AY(k^2)
\rightarrow 
Z_3-\frac32 Z_4 \, g^2_{\s R} C_A Y_{\s R}(k^2)
\label{renZ}
\end{align}
must be implemented on its r.h.s., 
with $Z_3$ and $Z_4$ the 
renormalization constants of the 
$Q^3$ and $Q^4$ vertices, respectively.

The presence of $Z_3$ and $Z_4$ 
converts the computation of the rhs of 
\1eq{DSEmass} into a highly nontrivial 
exercise, which requires, among other things, 
the detailed knowledge of the structure 
of the $Q^3$ and $Q^4$ vertices. Therefore,
as is common in this type of analysis,
we will simplify the situation by setting 
$Z_3=Z_4=1$.

Then, substituting \1eq{fixc}
 into~\1eq{DSEmass}, we 
obtain a second order algebraic equation for $\alpha_s$, given by
\begin{align}
    A\alpha_s^2+B\alpha_s+C=0,
    \label{quadraticmass}
\end{align}
where, passing to Euclidean space and using spherical coordinates,
\begin{align}
    A&=\frac{3C^2_A}{32\pi^3}F(0)\hspace{-0.1cm}\int_0^\infty\hspace{-0.2cm}\diff y\, y^2{\Delta}^2(y) Y(y) S'(y),\nonumber \\
    B&=-\frac{3C_A}{8\pi} F(0)\hspace{-0.1cm}\int_0^\infty\hspace{-0.2cm}\diff y\,  y^2{\Delta}^2(y) S'(y),\nonumber \\
    C&=-\int_0^\infty\hspace{-0.2cm}\diff y\,S'(y).
	\label{ABC}
\end{align}
with $A>0$ and $B,C<0$.
The unique positive solution of~\1eq{quadraticmass} is given by 
\begin{align}
    \aDSE=\frac{-B+\sqrt{B^2-4AC}}{2A},
    \label{aDSE}
\end{align}    
which shows how the existence of a positive coupling relies on a delicate interplay between the strength of the one- and two-loop dressed contributions in the gluon SDE.

We will now perform a numerical analysis in order to establish if the equality $\aDSE=\aBSE$ can be indeed realized, and, if so, at what value of the strong coupling $\alpha_s$. 

In order to fully appreciate the importance of employing a 
nontrivial $\fQ$
in this context, let us set $\fQ=1$ both in \1eq{defY} and 
\1eq{masslessBSE}. 
Then, a straightforward calculation yields 
the rather disparate set of values $\aDSE=0.42$
and  $\aDSE=0.27$. As we will see, the effect of 
using a physically motivated $\fQ$ will be a slight  
increase in $\aDSE$ combined with a considerable   
increase in $\aBSE$.
 
 Let us choose for $\fQ $ a fit to the data of~\fig{fig:fQ3}, given by~\cite{Athenodorou:2016oyh}
\begin{align}
    \fQ (q^2) &= \lambda\left[1 + b\ln\frac{q^2 + {\cal M}^2}{\mu^2}+c\ln\frac{q^2}{\mu^2}+\right.\nonumber \\&\left.+
    e\frac{{\cal M}^2 (q^2-\mu^2)}{(q^2+{\cal M}^2)(\mu^2+{\cal M}^2)} \right],
\label{vertex3}
\end{align}
with $\mu=4.3$ GeV the renormalization scale.  
We set $b=e=-5.30$, $c=5.40$, ${\cal M}=0.124$ GeV, 
but leave the scale factor $\lambda$ undetermined for the moment.

Next, using the same three-gluon vertex approximation~\noeq{vertf},~\1eq{defY} yields (in spherical coordinates\footnote{Here we set $t=\ell^2$ and $u=(k+\ell)^2$ and $k\!\cdot\!\ell=\sqrt{yt}\cos\omega$.} and $d=4$)
\begin{align}
    iY(y)&=\frac1{24\pi^3}\hspace{-0.1cm}\int_0^\infty\hspace{-0.2cm}\diff t\hspace{-0.1cm}\int_0^\pi\hspace{-0.2cm}\diff\omega\sin^2\omega \left[5+(y+\sqrt{yt}\cos\omega)/u
    \right]\nonumber \\
    &\times \fQ (u)\Delta(t)\Delta(u).
\end{align}
We emphasize that $Y$ is computed for the first time using full gluon propagators and a nonperturbative Ansatz for the three-gluon vertex; this is a 
major improvement, given that all previous treatments of this quantity 
were purely perturbative (one loop)~\cite{Binosi:2012sj}. 

\begin{figure}[!t]
    \centering
    \mbox{}\hspace{-0.8cm}
    \includegraphics[scale=0.6]{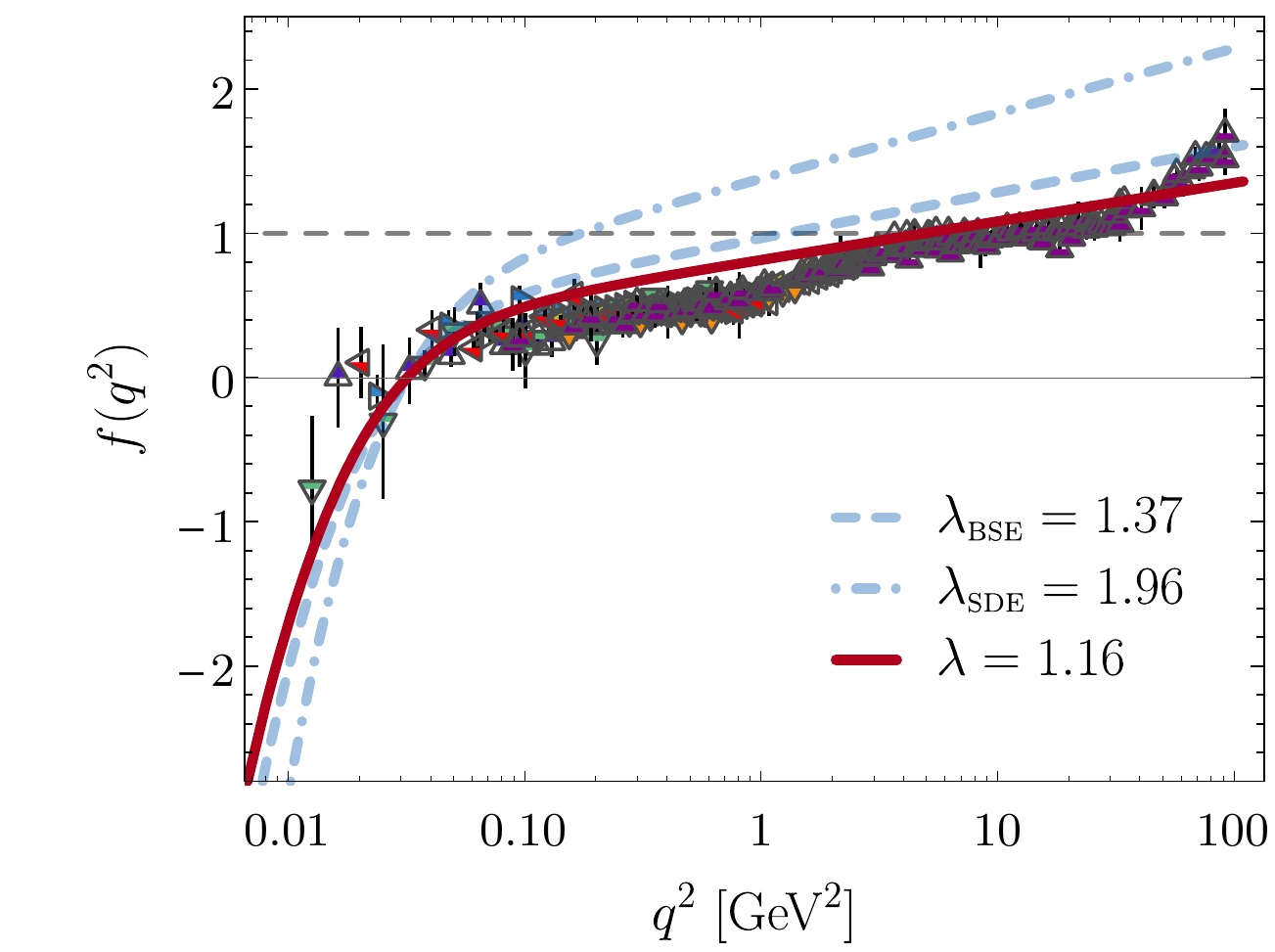}
    \caption{\label{fig:f} The three-gluon vertex form factor $\fQ(q^2)$ at $\lambda=1.16$ which leads to the equality $\aDSE=\aBSE=0.45$ (red continuous curve). The dashed (light blue) lines show the different  form factors needed in the BSE (dashed) and SDE (dot-dashed) to force the equality at $\aDSE=\aBSE=0.32$.}
\end{figure}

We then proceed as follows. To begin with, 
both in the evaluation of~\noeq{masslessBSE} and~\1eq{aDSE} 
we use as input for $\Delta(k^2)$ and $F(0)$ 
the lattice data of~\cite{Bogolubsky:2009dc}.
Then, we set in~\1eq{vertex3} the convenient starting value 
$\lambda_0=1$, and determine the value of the 
coupling $\aBSE = \alpha_0$ 
for which the BSE~\noeq{masslessBSE} yields the nontrivial solution $S'$;
specifically, we find that $\alpha_0 = 0.61$. 
Next, we substitute  $S'$ into~\1eq{ABC}
and compute the coefficients $A_0$, $B_0$ and $C_0$ of~\1eq{ABC}, whose values are (all in GeV$^2$) $A_0=156.2$, $B_0=-40.7$ and $C_0=-18.5$. 
Substituting them into~\1eq{aDSE}, one 
obtains $\aDSE=0.5$; evidently,  
$\aDSE \neq \alpha_0$. 
In order to achieve the desired equality $\aDSE=\aBSE$, note that,  
if $\lambda$ is moved from $\lambda_0=1$, the BSE will yield precisely the 
same solution as before provided that its coupling 
is rescaled to $\aBSE = \alpha_0/\lambda^2$ 
(recall that the BSE is quadratic in $\fQ$). 
In addition, since $Y$ is linear in $\fQ$, we will simply have that 
$A \to  \lambda A_0$, while $B$ and $C$ remain at their initial values.
Therefore, imposing the condition $\aDSE=\aBSE$ implies that the scale factor $\lambda$ has to be such that
\begin{align}
    \frac{-B_0+\sqrt{B_0^2-4\lambda A_0C_0}}{2\lambda A_0 } = \frac{\alpha_0}{\lambda^2},
\end{align}
or, equivalently, 
\begin{align}
C_0 \lambda^3 + \alpha_0 B_0 \lambda +\alpha^2_0 A_0 =0,   
\end{align}
whose only real solution is $\lambda \approx 1.16$.
A shown in in~\fig{fig:f}, 
the $\fQ$ obtained from \1eq{vertex3} using this special  value for $\lambda$
fits particularly well the lattice data.
Thus, the two couplings converge to the single value  $\aBSE=\aDSE=0.45$,
corresponding to $g=2.4$ at $\mu^2=4.3$ GeV, which is 20\% off 
the value used for $g$ in the lattice simulations mentioned above. 
 
As a final possibility, let us assign to the SDE and the BSE  
different forms of $\fQ$, by setting 
$\lBSE \neq \lDSE$.
This difference may be considered as a simple way of accounting for 
the fact that, while in the SDE all arguments of $\fQ(x,y,z)$ 
are integrated over (being virtual), in the BSE the third argument is associated with the external momentum $p$; this, in turn, may modify slightly the 
corresponding integrated strengths. 
Then, a straightforward repetition of the iteration procedure 
described above reveals that one may obtain $\aBSE=\aDSE=0.32$ by choosing 
$\lBSE = 1.37$ and $\lDSE=1.96$; 
the corresponding $\fQ$ are shown in~\fig{fig:f}. 

\section{Conclusions}

We have carried out an extensive  analysis of the interlocked
dynamics between the SDE of the gluon propagator $\Delta(q^2)$
and a BSE that generates massless bound state poles.
These poles constitute an indispensable ingredient  of the 
particular realization of the Schwinger 
mechanism employed in a series of works
in order to obtain infrared finite (massive)  
solutions for $\Delta(q^2)$. The notion of
coupling the two equations is novel, and its possible
ramifications for the overall self-consistency of the
entire formalism have not 
been explored before in the relevant literature. 

Our three main results may be summarized as follows. 
First, we have obtained a running gluon mass,  
displaying all expected physical features,  
directly from the solution of the BSE. This possibility
was envisaged in earlier works~\cite{Ibanez:2012zk}, but the two conditions discussed after \1eq{C1vsmass}, which are crucial 
for obtaining a positive-definite and monotonically decreasing
gluon mass,  were not fully appreciated. 
Second, we have carried out 
a nonperturbative computation of the  quantity $Y$, whose role is crucial 
for obtaining from the SDE a positive-definite gluon mass.
Third, we have demonstrated 
that the inclusion of the three-gluon vertex
is of paramount importance for the 
fulfillment of a basic 
self-consistency requirement.
In particular, the nontrivial
infrared dynamics of this vertex
compensate the original discrepancy in the
value of $\alpha_s$ used in the SDE and the BSE sectors, allowing finally for a single common value,  $\alpha_s=0.45$.

The deviation from the 
$\alpha_s=0.32$ estimated 
from the lattice simulations of~\cite{Athenodorou:2016oyh,Boucaud:2017obn} 
may be attributed to a variety of reasons. 

To begin with, 
the skeleton expansion of the BSE kernel has been 
truncated at the lowest order, shown in \fig{fig:bse-4g}. 
It would be very important to verify the impact of the 
next order corrections (``one-loop'' fully dressed).
In fact, even the impact of graph ($b_1$), whose vanishing seems to be an 
accident of setting the  
four-gluon vertex at tree level,
ought to be reconsidered, using a more complete 
structure for this vertex~\cite{Binosi:2014kka,Cyrol:2014kca}.

In addition, the transition from \1eq{renZ} to \1eq{quadraticmass} 
was implemented by setting into the former $Z_3=Z_4=1$.
A more complete treatment of this issue has been given in~\cite{Aguilar:2014tka};
the resulting kernel, however, is substantially more difficult to calculate,
and only Ans\"atze have been studied thus far. Unfortunately, 
the complicated nature of this problem makes progress in this 
direction rather slow.

Turning to $\fQ$, it is clear that the form 
of \1eq{vertf} is rather restrictive, given that the full tensorial 
basis for expanding $\Gamma$ consists of 14 elements.
In addition, $\fQ$ has been considered to be a function of 
a single variable (symmetric configuration : $q^2=r^2=p^2$). 
Clearly, a more complete integration over all available momenta
and angles could shift the coincidence value of $\alpha_s$ 
closer to $\alpha_s=0.32$, as exemplified in the last part of 
Section~5 by employing 
$\lambda\s{\mathrm{BSE}} \neq \lambda\s{\mathrm{DSE}}$.

Last but not least, the assumption that 
only the vertex ${\widetilde \Gamma}$ develops a 
massless pole may have to be revisited, allowing 
the remaining vertices, and especially the ghost-gluon vertex, to form part of a more complex BSE system.

We hope to return to some of the issues mentioned above 
in the near future.

\acknowledgments

We thank J. Rodriguez-Quintero for furnishing the lattice data of~\fig{fig:glprop}. The research of J. P. is supported by the Spanish MEYC under FPA2014-53631-C2-1-P and SEV-2014-0398, and Generalitat Valenciana under grant Prometeo II/2014/066.


%

\end{document}